\documentclass{article}
\usepackage[T1]{fontenc} 
\usepackage[utf8]{inputenc} 
\usepackage{ismir,amsmath,cite,url}
\usepackage{graphicx}
\usepackage{color}
\usepackage{enumitem}
\usepackage{bbm}
\usepackage{booktabs} 
\usepackage{subcaption}

\usepackage{lineno}

\title{Contrastive Learning of Musical Representations}

\oneauthor
{Janne Spijkervet \hfil John Ashley Burgoyne}
{Institute for Logic, Language, and Computation, University of Amsterdam \\ {\tt janne.spijkervet@gmail.com, j.a.burgoyne@uva.nl}}

\def\authorname{J. Spijkervet and J.A. Burgoyne}

\usepackage[bookmarks=false,pdfauthor={\authorname},pdfsubject={\papersubject},hidelinks]{hyperref}

\begin{document}

\maketitle

\begin{abstract}
While deep learning has enabled great advances in many areas of music, labeled music datasets remain especially hard, expensive, and time-consuming to create.
In this work, we introduce SimCLR to the music domain and contribute a large chain of audio data augmentations to form a simple framework for self-supervised, contrastive learning of musical representations: CLMR.
This approach works on raw time-domain music data and requires no labels to learn useful representations.
We evaluate CLMR in the downstream task of music classification on the MagnaTagATune and Million Song datasets and present an ablation study to test which of our music-related innovations over SimCLR are most effective.
A linear classifier trained on the proposed representations achieves a higher average precision than supervised models on the MagnaTagATune dataset, and performs comparably on the Million Song dataset.
Moreover, we show that CLMR's representations are transferable using out-of-domain datasets, indicating that our method has strong generalisability in music classification.
Lastly, we show that the proposed method allows data-efficient learning on smaller labeled datasets: we achieve an average precision of 33.1\% despite using only 259 labeled songs in the MagnaTagATune dataset (1\% of the full dataset) during linear evaluation.
To foster reproducibility and future research on self-supervised learning in music, we publicly release the pre-trained models and the source code of all experiments of this paper.
\end{abstract}
\section{Introduction}\label{sec:introduction}
Supervised learning methods have been widely used in musical tasks like chord recognition \cite{korzeniowski_fully_2016, chen_harmony_2019}, key detection \cite{korzeniowski_end--end_2017}, beat tracking \cite{bock_joint_2016}, music audio tagging \cite{pons_end--end_2017} and music recommendation \cite{van_den_oord_deep_2013}.
These methods require labeled corpora, which are difficult, expensive and time-consuming to create for music in particular \cite{doi:10.1080/09298215.2019.1613436}, while raw unlabeled music data is available in vast quantities.
Unsupervised alternatives to end-to-end deep learning for music are compelling, especially if they can generalise to smaller datasets.

Despite the importance of unsupervised learning for raw audio signals, unsupervised learning for musical tasks has yet to see breakthroughs comparable to those in supervised learning.
There have been successes with methods like PCA, PMSC's and spherical $k$-means that rely on a transformation pipeline \cite{hamel2011temporal, dieleman_feature_learning}, and very recently with self-supervised methods in the time-frequency domain for general audio classifiation tasks\cite{cramer2019,tagliasacchi2020,pmlr-v130-al-tahan21a,saeed2021}, but learning effective representations of raw audio in an unsupervised manner has remained elusive for musical tasks.

\begin{figure}[t]
    \includegraphics[width=\columnwidth]{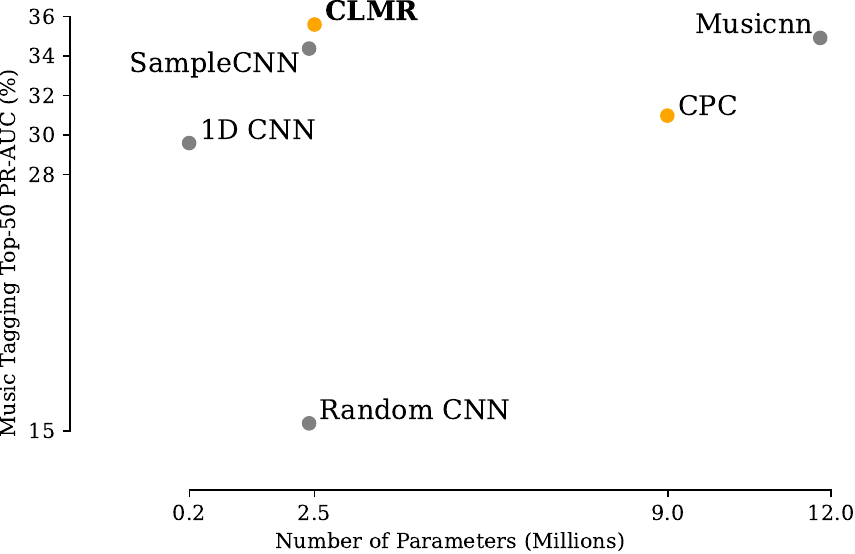}
    \caption{Performance and model complexity comparison of supervised models (grey) and self-supervised models (ours) in music classification of raw audio waveforms on the Magna\-Tag\-A\-Tune dataset to evaluate musical representations.
Supervised models were trained end-to-end, while CLMR and CPC are pre-trained without ground truth: their scores are obtained by training a \textit{linear} classifier on their learned representations but nonetheless perform competitively to the supervised models.}
    \label{fig:music_tagging_overview}
\end{figure}

Self-supervised representation learning is an unsupervised learning paradigm that has demonstrated advances across many tasks and research domains \cite{dosovitskiy2015discriminative, oord_representation_2019, hjelm_learning_2019,chen_simple_2020,grill2020bootstrap}.
This includes the ability to use substantially less labeled data when fine-tuning on a specific task \cite{henaff2019data,chen_simple_2020,chen2020big}.
Without ground truth, there can be no ordinary loss function for training; self-supervised learning trains by way of a proxy loss function instead.
One way to preserve the amount of useful information during self-supervised learning is to define the proxy loss function with respect to a relatively simple pretext task, with the idea that a representation that is good for the pretext task will also be useful for downstream tasks.
Many approaches rely on heuristics to design pretext tasks \cite{doersch_unsupervised_2015,zhang2016colorful}, e.g., by witholding a pitch transformation \cite{spice}.
Alternatively, \emph{contrastive representation learning} formulates the proxy loss directly on the learned representations and relies on contrasting multiple, slightly differing versions of any one example by often using negative sampling strategies \cite{tian2019contrastive,he2019moco,chen_simple_2020} or by bootstrapping the representations \cite{grill2020bootstrap}.

In this paper, we combine the insights of a simple contrastive learning framework for images, SimCLR \cite{chen_simple_2020}, with recent advances in representation learning for audio in the time domain \cite{lee2018samplecnn}. We also contribute a pipeline of data augmentations on musical audio, to form a simple framework for self-supervised, contrastive learning of representations of raw waveforms of music.
To compare the effectiveness of this simple framework compared to a more complex self-supervised learning objective, we also evaluate representations learned by contrastive predictive coding (CPC) \cite{oord_representation_2019}.
The self-supervised models are evaluated on the downstream music tagging task, enabling us to evaluate their versatility: music tags describe many characteristics of music, e.g., genre, instrumentation and dynamics.
Our key contributions are the following.
\begin{itemize}[topsep=0pt, partopsep=0pt, leftmargin=13pt, parsep=0pt, itemsep=4pt]
    \item CLMR achieves strong performance on the music classification task compared to supervised models, despite self-supervised pre-training and training a linear classifier on the downstream task with raw signals of musical audio (see Figure~\ref{fig:music_tagging_overview}).
    \item CLMR enables efficient classification: we achieve comparable performance using as few as 1\% of the labeled data.
    \item We show the out-of-domain transferability of representations learned from pre-training CLMR on entirely different corpora of musical audio.
    \item CLMR can learn from \emph{any} dataset of raw music audio, requiring neither transformations nor fine-tuning on the input data; nor do the models require manually annotated labels for pre-training.
    \item We provide an ablation study on the effectiveness of individual audio data augmentations.
\end{itemize}

\section{Related Work}
\label{sec:related}

The goal of representation learning is to identify features that make  prediction tasks easier and more robust to the complex variations of natural data \cite{bengio2013representation}.
In unsupervised representation learning, generative modeling and likelihood-based models typically find useful representations of the data by attempting to reconstruct the observations on the basis of their learned representations \cite{goodfellow2014generative, unsupervised_gan}.
\emph{Self-supervised} representation learning aims to identify the explanatory factors of the data using an objective that is formulated with respect to the learned representations directly \cite{doersch_unsupervised_2015,zhang2016colorful,oord_representation_2019,henaff2019data,grill2020bootstrap}.

Compared to vision, work on self-supervised learning in audio is still very limited, but there are a number of works that appeared very recently.
Contrastive predictive coding is a universal approach to contrastive learning, and has been successful for speaker and phoneme classification using raw audio, among other tasks \cite{oord_representation_2019}. PASE \cite{Pascual2019} introduces several self-supervised workers that solve regression or binary discrimation tasks, that jointly optimise an encoder for speech recognition. To improve the representations for mismatched acoustic conditions and their transferability, they apply augmentations to the input speech signal \cite{pase_plus}.
In music information retrieval, recent advances have been made in self-supervised pitch estimation \cite{spice}, closely matching supervised, state-of-the-art baselines \cite{crepe} despite being trained without ground truth labels. $L^3$-Net learns deep embeddings from audio-visual correspondence in videos by way of self-supervised learning \cite{cramer2019}. Their work uses mel-spectrograms for audio and requires more than 40 million audio-video training samples to learn optimal embeddings. Audio2Vec also operates in the time-frequency domain and learns by reconstructing spectrogram slices from past and future slices \cite{tagliasacchi2020}. With limited data, Audio2Vec outperforms supervised models in pitch and instrument classification. CLAR also uses a contrastive learning objective, and computes a loss on a concatenation of representations learned from both raw audio and mel-spectrograms\cite{pmlr-v130-al-tahan21a}. COLA uses a similar method with mel-spectrograms only, and uses bilinear comparisons instead of cosine similarity \cite{saeed2021}. Both works are evaluated on speech command, environmental sound classification, and on pitch and instrument classification on the NSynth dataset\cite{nsynth2017}.

\section{Method}
\label{sec:method}
This work builds on SimCLR, a simple contrastive learning framework of visual representations \cite{chen_simple_2020}. Despite a task-agnostic, labelless discriminative pre-training approach, a linear classifier achieved performance comparable to fully supervised models in many image classification benchmarks.
Its learning objective is to maximise the agreement of latent representations of augmented views of the same image using a contrastive loss.
In Section~\ref{sec:related}, we will continue an overview of contrastive learning.

In CLMR, we adapt this framework to the domain of raw music audio. While most core components of CLMR have appeared in previous work, its ability to model waveforms of music cannot be explained by a single design choice, but by their composition.
We will first elaborate the four core components in the following subsections:
\begin{itemize}[topsep=0pt, partopsep=0pt, leftmargin=13pt, parsep=0pt, itemsep=4pt]
    \item A stochastic composition of data augmentations that produces two correlated, augmented examples of the same audio fragment, the `positive pair', denoted as $x_i$ and $x_j$.
    \item An encoder neural network $g_{\text{enc}}(\cdot)$ that maps the augmented examples to their latent representations.
    \item A projector neural network $g_{\text{proj}}(\cdot)$ that maps the encoded representations to the latent space where the contrastive loss is formulated.
    \item A contrastive loss function, which aims to identify $x_j$ from the negative examples in the batch $\{x_{k\neq i}\}$ for a given $x_i$.
\end{itemize}

\begin{figure}[t]
    \includegraphics[width=\columnwidth]{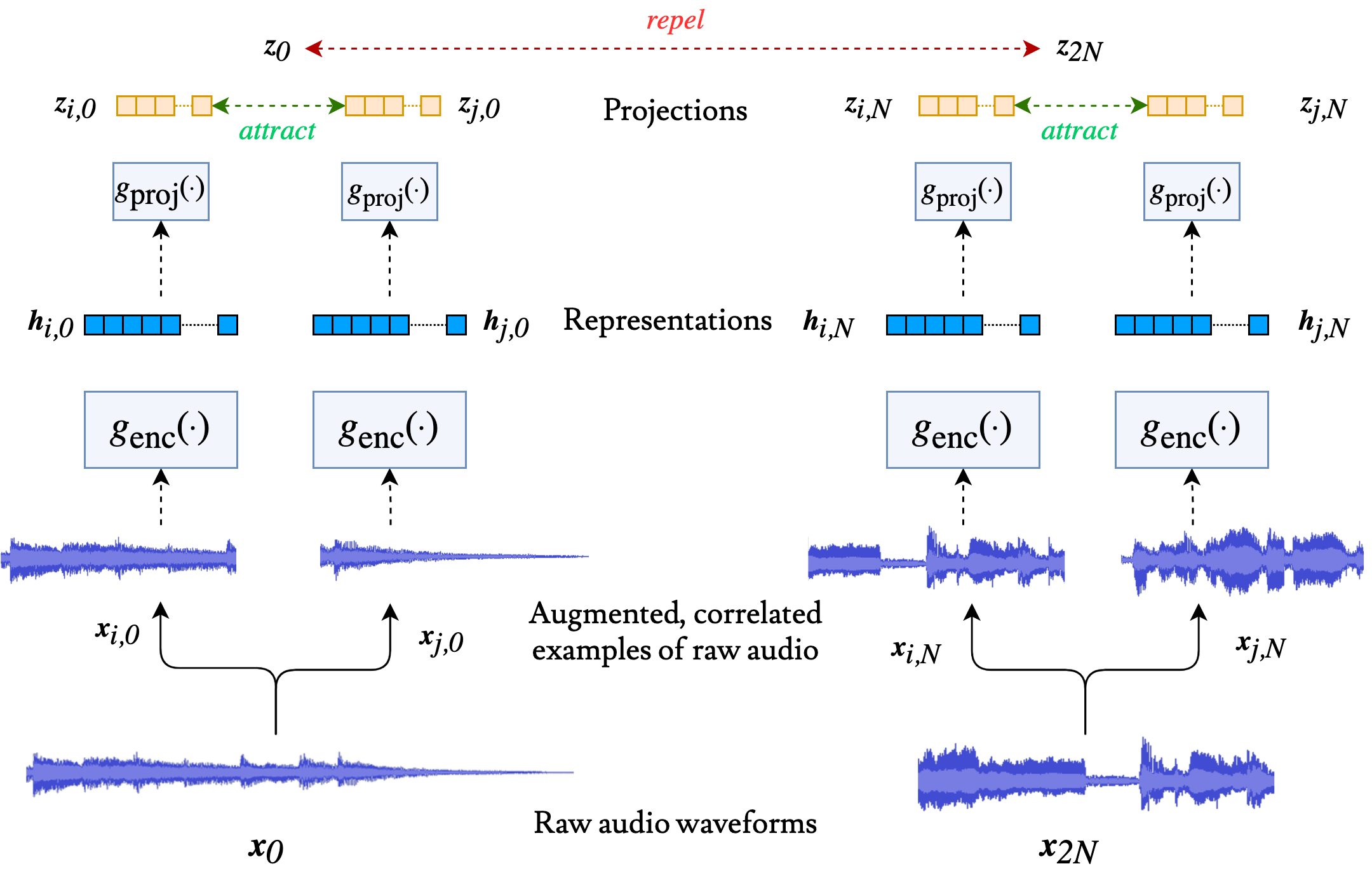}
    \caption{The complete framework operating on raw audio, in which the contrastive learning objective is directly formulated in the latent space of correlated, augmented examples of pairs of raw audio waveforms of music.}
    \label{fig:clmr_model}
\end{figure}

The complete framework is visualised in Figure~\ref{fig:clmr_model}.

\subsection{Data Augmentations}
We designed a comprehensive chain of audio augmentations for raw audio waveforms of music to make it harder for the model to identify the correct pair of examples. For details, see Appendix \ref{appendix:data_augmentation}\footnote{The supplementary material can be found at the accompanying webpage of this paper : \url{https://spijkervet.github.io/CLMR}}.
Each consecutive augmentation is stochastically applied on ${x_i}$ and ${x_j}$ independently, i.e., each augmentation has an independent probability $p_{\text{transform}}$ of being applied to the audio.
The order of augmentations applied to audio is carefully considered, e.g., applying a delay effect \textit{after} reverberation empirically gives an entirely different result in music.

\begin{enumerate}[topsep=0pt, partopsep=0pt, leftmargin=13pt, parsep=0pt, itemsep=4pt]
    \item A random fragment of size $s$ is selected from a piece of music, without trimming silence (e.g., the intro or outro of a song).
    The two examples $x_i$ and $x_j$ from the same audio fragment can overlap or be very disjoint, allowing the model to infer both local and global structures.
    \item The polarity of the audio signal is inverted, i.e., the amplitude is multiplied by $-1$.
    \item Additive white Gaussian noise is added with a signal-to-noise ratio of 80 decibels to the original signal.
    \item The gain is reduced between $[-6, 0]$ decibels.
    \item A frequency filter is applied to the signal.
    A coin flip determines whether it is a low-pass or a high-pass filter. The cut-off frequencies are drawn from uniform distributions on $[2200, 4000]$ or $[200,1200]$ Hz respectively.
    \item The signal is delayed and added to the original signal with a volume factor of 0.5.
    The delay is randomly sampled between 200-500ms, in 50ms increments.
    \item The signal is pitch shifted.
    The pitch transposition interval is drawn from a uniform distribution of semitones between $[-5, 5]$, i.e., a perfect fourth compared to the original signal's scale.
    \item Reverb is added to alter the signal's acoustics.
    The impulse response's room size, reverbation and damping factor is drawn from a uniform distribution on $[0, 100]$.
\end{enumerate}
\medskip
The space of augmentations is not limited to these operations and could be extended to, e.g., randomly applying chorus, distortion and other modulations. Some of these have been shown to improve performance in self-supervised learning for automatic speech recognition in the time-domain as well \cite{pase_plus, wavaugment2020}.


\subsection{Batch Composition}
A larger batch size $N$ makes the contrastive learning objective harder -- there are simply more negative examples the anchor sample needs to identify the positive sample from -- but it can substantially improve model performance \cite{chen_simple_2020}.
We sample one song from the batch, augment it into two examples, and treat them as the positive pair.
We treated the remaining $2(N-1)$ examples in the batch as negative examples, and did not sample the negative examples explicitly.
Larger batch sizes introduces a practical problem for raw audio when training on a GPU, as their input dimensionality increases for higher sample rates.
When training on multiple GPU's, we used global batch normalisation, i.e., we aggregate the batch statistics over all devices during parallel training, to avoid potential leakage of batch statistics because the positive examples are sampled on the same device (which improves training loss, but counteracts learning of useful representations).

\subsection{Encoder}
To directly compare a state-of-the-art end-to-end supervised model used in music classification on raw waveforms against a self-supervised model, we use the SampleCNN architecture as our encoder \cite{lee2018samplecnn}.
Similarly, we use a fixed audio input of 59\,049 samples with a sample rate of 22\,050~Hz.
In this configuration, the SampleCNN encoder $g_{\text{enc}}$ consists of 9 one-dimensional convolution blocks, each with a filter size of 3, batch normalisation, ReLU activation and max pooling with pool size 3.
The final output layer is removed, which yields a 512-dimensional feature vector $h_i$ for every audio input.
The feature vectors from the encoder can be directly used in the learning objective, but formulating the objective on encodings mapped to a different latent space by a parameterised function helps the effectiveness of the representations \cite{chen_simple_2020}.
In our experiments, we use a non-linear layer $z_i = W^{(2)}\operatorname{ReLU}(W^{(1)}h_i)$ with an output dimensionality of 128 as the projection head $g_{\text{proj}}$.
There are 2.5 million trainable parameters in total, which is put in comparison with other state-of-the-art models in Figure~\ref{fig:music_tagging_overview}.

We used 96 examples per batch and the aforedescribed encoder configuration to directly compare our self-supervised performance with the equally expressive fully supervised method \cite{lee2018samplecnn}.
We ran experiments with batch sizes of 96 on 2$\times$ NVIDIA 1080Ti, while for larger batches up to 4 $\times$ Titan RTX's were used. With 2 1080Ti's, it takes $\sim$5 days to train 1\,000 epochs on our largest dataset.

\subsection{Contrastive Loss Function}
In keeping with recent findings on several objective functions in contrastive learning \cite{chen_simple_2020}, the contrastive loss function used in this model is normalised temperature-scaled cross-entropy loss, commonly denoted as \emph{NT-Xent loss}:
\begin{equation}
    \label{ntxent_loss}
    \ell_{i, j}=-\log \frac{\exp \left(\operatorname{sim}\left(z_{i}, z_{j}\right) / \tau\right)}{\sum_{k=1}^{2 N} \mathbbm{1}_{[k \neq i]} \exp \left(\operatorname{sim}\left(z_{i}, z_{k}\right) / \tau\right)}
\end{equation}

The pairwise similarity is measured using cosine similarity and the temperature parameter $\tau$ helps the model learn from hard negatives.
The indicator function $\mathbbm{1}_{[k \neq i]}$ evaluates to $1$ iff $k\neq i$.
This loss is computed for all pairs, both $(z_i, z_j)$ and $(z_j, z_i)$, for $i\neq j$. 


\subsection{Contrastive Predictive Coding}
We adjusted the original CPC encoder $g_{\text{enc}}$ \cite{oord_representation_2019} to a deeper architecture for more direct comparison \cite{lee2018samplecnn}. The encoder $g_{\text{enc}}$ consists of 7 layers with 512 filters each, and filter sizes $[10, 6, 4, 4, 4, 2, 2]$ and strides $[5, 3, 2, 2, 2, 2, 2]$.
Instead of relying on max-pooling, the filter sizes and strides are adjusted to parameterise and facilitate downsampling.
We also increased the number of prediction steps to 20, effectively asking the network to predict 100~ms of audio into the future. The batch size is set to 64 from which 15 negative examples in the contrastive loss are drawn.

\subsection{Linear Evaluation}
\label{sec:evaluation}
The evaluation of representations learned by self-supervised models is commonly done with linear evaluation \cite{oord_representation_2019,hjelm_learning_2019,chen_simple_2020}, which measures how linearly separable the relevant classes are under the learned representations.
We obtain the representations for all datapoints from a frozen CLMR network after pre-training has converged, and train a linear classifier using these self-supervised representations on the downstream task of music classification.
For CPC, the representations are extracted from the autoregressor, yielding a context vector of size $(20,256)$, which is global-average pooled to obtain a single vector of $512$ dimensions.
For CLMR, the last 512-dimensional vector $h$ from the encoder is used instead of $z$ from the projection head because that yielded consistently better results for all our experiments.
We compute the evaluation metrics on a held-out test set, averaged over three runs on the training set using different random seeds.

\subsection{Optimisers}
We use the Adam optimiser \cite{adam_optimizer} with a learning rate of $0.0003$ and $\beta_1 = 0.9$ and $\beta_2 = 0.999$ during pre-training and employ He initialisation for all convolutional layers. The temperature parameter $\tau$ is set to 0.5, since we observed consistent results regardless of varying batch sizes and temperature $\tau \in \{0.1, 0.5, 1.0\}$.
For linear evaluation, we use the Adam optimiser with a learning rate of $0.0003$ and a weight decay of $10^{-6}$. Backpropagation is only done in the final (linear) head for all experiments in this paper.
We also employ an early stopping mechanism when the validation scores do not improve for 5 epochs.

\section{Experimental Results}\label{sec:results}

\begin{table}[t]
    \centering
    \footnotesize
    \begin{tabular}{@{}llcc@{}}\toprule
        Model & Dataset & $\text{ROC-AUC}$ & $\text{PR-AUC}$ \\ \midrule
        CLMR (ours) & MTAT & 88.7 (\textbf{89.3}) & 35.6 (\textbf{36.0}) \\
        Musicnn\cite{pons_end--end_2017}$^\dagger$ & MTAT & 89.0 & 34.9 \\
        SampleCNN\cite{lee2018samplecnn}$^\dagger$ & MTAT & 88.6 & 34.4 \\
        CPC (ours) & MTAT & 86.6 (88.0) & 31.0 (33.0) \\
        1D CNN\cite{dieleman2014end}$^\dagger$ & MTAT & 85.6 & 29.6 \\\midrule
        Transformer\cite{won2021transformer}$^{\dagger\S}$ & MSD & \textbf{89.7} & \textbf{34.8} \\
        Musicnn\cite{pons_end--end_2017}$^\dagger$ & MSD & 88.0 & 28.7 \\
        SampleCNN\cite{lee2018samplecnn}$^\dagger$ & MSD & 87.9 & 28.5 \\
        CLMR (ours) & MSD & 85.7 & 25.0 \\
        \bottomrule
    \end{tabular}
    \caption{Tag prediction performance on the Magna\-Tag\-A\-Tune (MTAT) dataset and Million Song Dataset (MSD), compared with fully supervised models$^{(\dagger)}$ trained on raw audio waveforms.
    We omit most works that operate on (mel-) spectrograms$^{(\S)}$ to make a fair comparison with our approach on raw audio. For reference, we add the Transformer model that is the current state-of-the-art in music tagging.
For the self-supervised models, the scores are obtained by training a \emph{linear}, logistic regression classifier using the frozen representations from self-supervised pre-training.
Scores in brackets show performance when adding a hidden layer to the linear classifier.}
    \label{tab:results}
\end{table}

\subsection{Datasets}
We evaluated the quality of our representations with music classification experiments.
Predicting the top~50 semantic tags in the Magna\-Tag\-A\-Tune and Million Song datasets  \cite{law2009evaluation,Bertin-Mahieux2011} is a popular benchmark for music classification.
These semantic tags are annotated by human listeners, and have a
varying degree of abstraction and describe many facets of music,
including genre, instrumentation and dynamics.
It is a multi-label classification task: each track can have multiple tags, of which we use the 50 most frequently occuring to compare our performance against supervised benchmarks.

The Magna\-Tag\-A\-Tune dataset consists of 25k music clips from 6\,622 unique songs, of which we use about 187k fragments of 2.6 seconds for training, and the same train/test split as previous work \cite{pons_end--end_2017, lee2018samplecnn, dieleman_feature_learning}.
The Million Song Dataset contains a million songs, of which about 240k previews of 30 seconds are available and labeled with Last.FM tag annotations.
We only use the train, validation and test split of 201\,680 / 11\,774 / 28\,435 songs as used in previous work \cite{pons_end--end_2017, lee2018samplecnn}, not all million songs during self-supervised pre-training.
This results in 2.2 million music fragments of 2.6 seconds for training, i.e., almost 1\,600 hours of music.
The tags for the Million Song Dataset also contain overlapping genre and semantic tags, e.g., `beautiful', `happy' and `sad', which are arguably harder to separate during the linear evaluation phase.

We use average tag-wise area under the receiver operating characteristic curve  (ROC-AUC) and average precision (PR-AUC) scores as evaluation metrics.
They are measured globally for the whole dataset, i.e., for the tag metric we measure the retrieval performance on the tag dimension (column-wise) and for the clip metric we measure the performance on the clip dimension (row-wise).
PR-AUC is calculated in addition to ROC-AUC, because ROC-AUC scores can be over-optimistic for imbalanced datasets like Magna\-Tag\-A\-Tune \cite{David-2006}.

\subsection{Quantitative Evaluation}
The most important goal set out in this paper is to evaluate the difference in performance between an otherwise identical, fully supervised network when learning representations using a self-supervised objective.

CLMR exceeds the supervised benchmark for the Magna\-Tag\-A\-Tune dataset with a PR-AUC of 35.6\%, despite task-agnostic, self-supervised pre-training and a \textit{linear} classifier for training, as shown in Table~\ref{tab:results}.
An additional 0.4\% PR-AUC performance gain is added by adding an extra hidden layer to the classifier. When increasing the batch size and the number of parameters, we observe another performance gain to 37.0\% PR-AUC as show in Appendix \ref{appendix:exp_mini_batch_size}.
The performance on the larger Million Song Dataset is lower compared to the supervised benchmark, and especially to the current state-of-the-art model that is trained using mel-spectrograms\cite{won2021transformer}, but is still remarkable given the use of a linear classifier. The tags in the Million Song Dataset are semantically more complex, e.g., `catchy', `sexy', `happy', and have more similar genre tags, e.g., `progressive rock', `classic rock' and `indie rock', which our proposed contrastive learning method may not distinguish.

CPC also shows competitive performance with fully supervised models in the music classification task.
Despite CPC's good performance, self-supervised training does not require a memory bank or more complex loss functions, e.g., those incorporating mutual information or more explicit negative sampling strategies, to learn useful representations.

We also analyse the quality of our representations, showing they can cleanly separate audio fragments from different classes, and visualise the convolution filters of the self-supervised models in Appendix \ref{appendix:qualitative_results}.

\subsection{Data Augmentations}
\label{sec:data_augmentations}
The CLMR model relies on a pipeline of strong data augmentations to facilitate the learning of representations that are more robust and allow for better generalisation in the downstream task.
In Table~\ref{tab:transformation_study}, we show the linear
evaluation scores obtained by taking a random crop of audio and performing one additional, individual augmentation.
While all datasets contain songs of variable length, we always sample a random crop of audio of the same size before applying other augmentations.
This makes it harder to assess the individual contribution of each augmentation to the downstream task performance.
We therefore consider an asymmetric data transformation setting: we only apply the augmentation(s) to one branch of the framework, while we settle with an identity function for the other branch (i.e., $t(x_j) = x_j$) \cite{chen_simple_2020}.
The model is pre-trained from scratch for 1\,000 epochs after which linear evaluation is performed.

When only taking a random crop of audio, we achieve a PR-AUC score of $30.5$.
Most individual augmentations show an increase in performance, while adding gain or delay does not impact performance as much.
Adding a filter to the augmentation pipeline increases the downstream performance more significantly.

\begin{table}[t]
    \centering
    \footnotesize
    \begin{tabular}{lcccc}
    \toprule
    &\multicolumn{2}{c}{Tag} &\multicolumn{2}{c}{Clip}\\
    \cmidrule(lr){2-3} \cmidrule(lr){4-5}
    Transform & $\text{ROC-AUC}$ & $\text{PR-AUC}$ & $\text{ROC-AUC}$ & $\text{PR-AUC}$  \\
    \midrule
    Filter & 87.6 & 33.3 & 92.5 & 67.9 \\
    Reverb & 86.5 & 31.7 & 91.8 & 65.8 \\
    Polarity & 86.3 & 31.5 & 91.7 & 65.7 \\
    Noise & 86.1 & 31.5 & 91.5 & 65.5 \\
    Pitch & 86.4 & 31.5 & 91.5 & 65.3 \\
    Gain & 86.2 & 31.1 & 91.5 & 65.1 \\
    Delay & 85.8 & 30.5	& 91.3 & 64.9 \\
    Crop & 85.8 & 30.5 & 91.3 & 64.8  \\
    \bottomrule
    \end{tabular}
    \caption{CLMR music tagging performance using a random crop together with one other audio data augmentation.}
    \label{tab:transformation_study}
\end{table}

\begin{figure}[t]
    \centering
    \includegraphics[width=0.95\columnwidth]{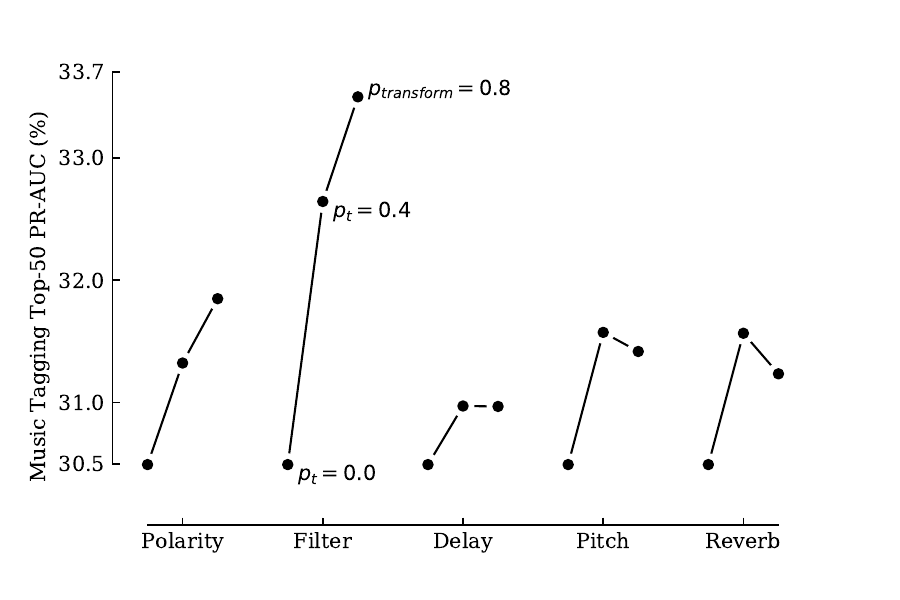}
    \caption{$\mathrm{PR-AUC}_{\mathrm{TAG}}$ scores for transformations under different, consecutive probabilities $p \in \{ 0.0, 0.4, 0.8 \}$}
    \label{fig:transformation_probabilities}
\end{figure}

Besides evaluating the individual contribution of each augmentation with augmentation probability $p_t = 1$, we also vary $p_t \in \{ 0, 0.4, 0.8 \}$.
This is done to assess the optimal amount of augmentation to each example, i.e., the contrastive learning task should neither be too hard, nor too simple, for learning effective representations for the music classification task.
The linear evaluation PR-AUC score is shown for each augmentation under a different probability $p_t$ in Figure~\ref{fig:transformation_probabilities}.
For the Polarity and Filter transformations, performing them more often with a probability of $p_t = 0.8$ is beneficial.
For the Delay, Pitch and Reverb transformations, a transformation probability of $p_t = 0.4$ works better than performing them more aggressively.
Generally, we find that strong data augmentations result in more robust representations and better downstream task performance.

\subsection{Data Efficient Classification Experiments}
\begin{figure}[t]
    \centering
    \includegraphics[width=0.95\columnwidth]{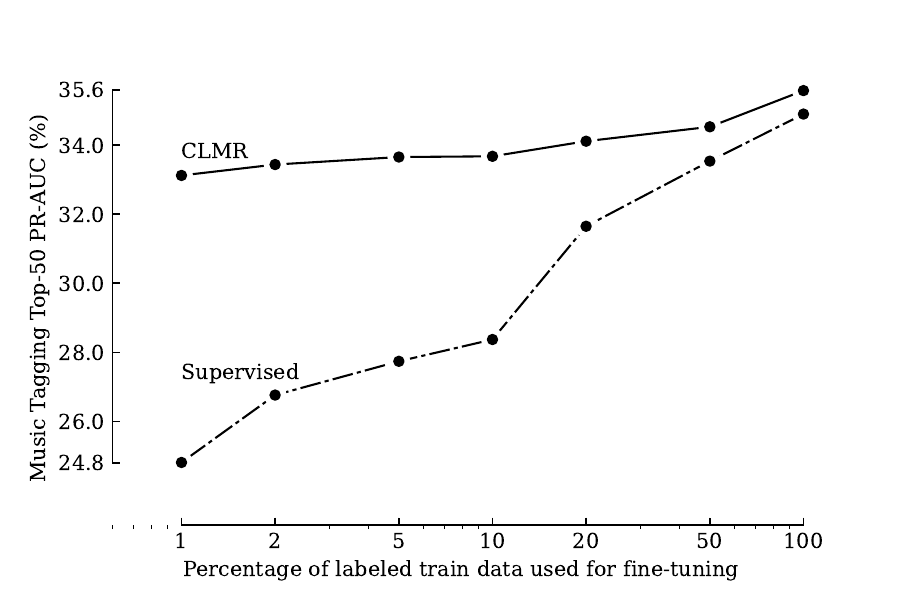}
    \caption{Percentage of labels used for training vs.
the achieved $\mathrm{PR-AUC}_{\mathrm{TAG}}$ score on the MTAT dataset.}
    \label{fig:perc_train_data_magnatagatune}
\end{figure}

\begin{figure}[t]
    \centering
    \includegraphics[width=0.95\columnwidth]{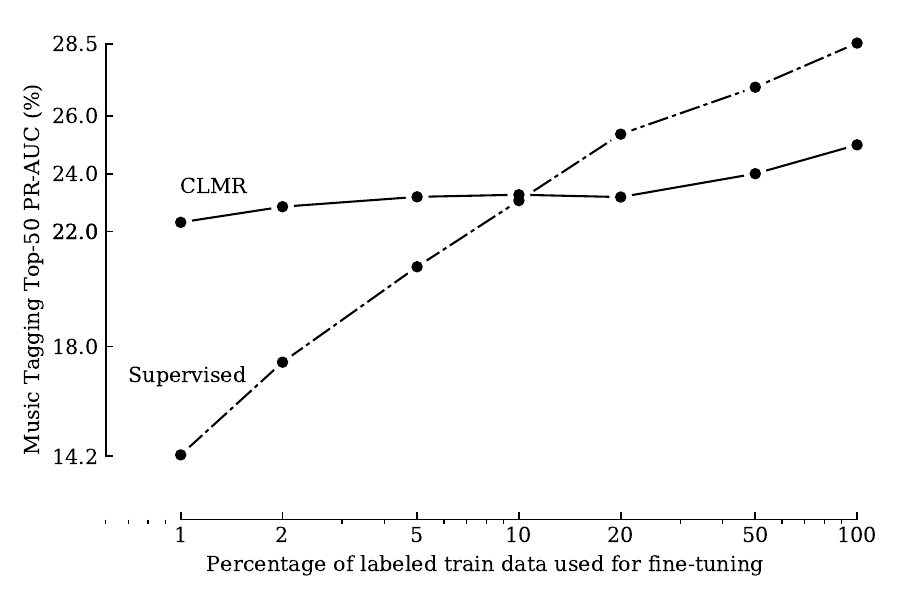}
    \caption{Percentage of labels used for training vs.
the achieved $\mathrm{PR-AUC}_{\mathrm{TAG}}$ score on the MSD.}
    \label{fig:perc_train_data_msd}
\end{figure}

To test the efficient classification capability of the CLMR model, we train the linear classifier on consecutive, class-balanced subsets of the labels in the train dataset and report its performance.
During the task-agnostic, self-supervised pre-training phase, 100\% of the data is used. Figures \ref{fig:perc_train_data_magnatagatune} and~\ref{fig:perc_train_data_msd} show the PR-AUC scores obtained when increasing the amount of labels available during training.
For both datasets, self-supervised pre-training greatly improves performance when less labeled data is available.
Using 100 times fewer labeled songs, i.e., only 259 songs, CLMR scores 33.1\% PR-AUC compared to 24.8\% PR-AUC obtained with an equivalent, end-to-end trained supervised model trained on about 25\,000 songs.
Pre-training using a self-supervised objective without labels therefore substantially improves efficient classification: only $1\%$ of the labels are required while maintaining a similar performance.
For the Million Song Dataset, a fully supervised model exceeds CLMR at $10\%$ of the labels, which are 24\,190 unique songs in total.

\subsection{Transfer Learning Experiments}
To test the out-of-domain generalisability of the learned representations, we pre-trained CLMR on entirely different music datasets.
After pre-training, we freeze the weights of the network, i.e., we do not fine-tune the encoder, and subsequently perform the linear evaluation procedure outlined in Section~\ref{sec:evaluation}.
While originally made for chord recognition, we use 461 contemporary pop songs recorded between the 1940's and 2000's from the McGill Billboard dataset \cite{burgoyne_billboard}.
The Free Music Archive dataset \cite{fma_dataset} consists of 22\,413 songs for the `medium' version, and the fault-filtered GTZAN dataset \cite{tzanetakis2002musical,sturm2013gtzan} contains 930 fragments of 30 seconds, both popular for music classification.

The results of the transfer learning experiments are shown in Table~\ref{tab:magnatagatune_results}.
Both CPC and CLMR show the ability to learn effective representations from out-of-domain datasets without ground truth, and even exceed accuracy scores of previous, supervised end-to-end systems on raw audio \cite{dieleman2014end}.
Moreover, both models even demonstrate the ability to learn useful representations on the much smaller GTZAN and Billboard datasets.
The CLMR model performs better when it is pre-trained on larger datasets, which is expected as it heavily relies on the number of unique, independent examples that make the contrastive learning task harder, resulting in more robust representations.
When pre-training on smaller datasets, CPC can find more useful representations, especially when adding an extra hidden layer to the fine-tune head.

\begin{table}[t]
    \centering
    \footnotesize
    \begin{tabular}{lccc}
    \toprule
    Model & Train Dataset & $\text{ROC-AUC}_\text{TAG}$ & $\text{PR-AUC}_\text{TAG}$ \\
    \midrule
    CLMR & MSD &  \textbf{87.8} & \textbf{33.1} \\  
    CPC & FMA & 86.3 (87.8) & 30.7 (32.5) \\
    CLMR & FMA & 86.2 (86.6) & 30.6 (31.2) \\
    CPC & Billboard & 85.8 (86.3) & 29.7 (30.2) \\
    CPC & GTZAN & 83.4 (86.0) & 26.9 (29.7) \\
    CLMR & Billboard & 82.7 (84.2) & 26.9 (27.8) \\
    CLMR & GTZAN & 81.9 (85.4) & 26.2 (29.5) \\
    \bottomrule
    \end{tabular}
    \caption{Transfer learning experiments for CLMR and CPC, which are trained on a separate dataset and evaluated on the Magna\-Tag\-A\-Tune dataset.
The reported scores are obtained with a frozen, pre-trained encoder and a linear classifier. Scores in parenthesis are obtained when adding one extra hidden layer to the classifier.}
    \label{tab:magnatagatune_results}
\end{table}

\section{Conclusion}\label{sec:conclusion}
In this paper, we presented CLMR, a self-supervised contrastive learning framework that learns useful representations of raw waveforms of musical audio.
The framework requires no preprocessing of the input audio and is trained without ground truth, which enables simple and straightforward pre-training on music datasets of unprecedented scale.
We tested the learned, task-agnostic representations by training a linear classifier on the music classification task on the Magna\-Tag\-A\-Tune and Million Song datasets, achieving competitive performance compared to fully supervised models.
We also showed that CLMR can achieve comparable performance using 100 times fewer labeled songs, and demonstrated the out-of-domain transferability of representations learned from pre-training on entirely different datasets of music.
To foster reproducibility and future research on self-supervised learning in music information retrieval, we publicly release the pre-trained models and the source code of all experiments of this paper\footnote{\url{https://github.com/spijkervet/CLMR}}.
The simplicity of training the model without any labels and without preprocessing the audio, together with encouraging results obtained with a single linear layer optimised for a challenging music task, are exciting developments towards unsupervised learning on raw musical audio.

\section{Acknowledgements}
We would like to thank Jordan B.L. Smith, Wilker Aziz and Keunwoo Choi for their feedback on the draft. We would also like to extend our gratitude to the University of Amsterdam and SURFsara for giving us access to their Research Capacity Computing Services GPU cluster.

\bibliography{main}

\cleardoublepage
\appendix
\counterwithin{figure}{section}
\counterwithin{table}{section}
\onecolumn

\appendix
\begin{center}
    Supplementary Material for Contrastive Learning of Musical Representations
\end{center}

\section{Audio Preprocessing}
In this paper, we used raw audio waveform data for training in both the pre-training and linear evaluation phases. The default audio sample rate for all experiments is 22\,050~Hz, except for the sample rate experiment in Section \ref{appendix:sample_rates}. The Magna\-Tag\-A\-Tune dataset contains monophonic 30-second audio fragments in MP3 format, sampled at 16\,000~Hz. Some of the audio fragments originate from the same song. We reconstructed the original song by concatenating the fragments into a single file, to avoid occurances of fragments of the same song in the same batch of positive- and negative pairs, thereby ensuring i.i.d. data for training.

The audio files from the Million Song Dataset were obtained from the 7digital service, which provides stereo 30-second audio fragments in MP3 format sampled at 44\,100~Hz.

All files were re-sampled to 22\,000~Hz, 16\,000~Hz and 8\,000~Hz and decoded to the PCM format with $\texttt{ffmpeg}$, using the following command:
\begin{verbatim}
    ffmpeg -i {input_file}.mp3 -ar {target_sample_rate} {output_file}.wav
\end{verbatim}

This is the only preprocessing step that we performed before training.

\section{Data Augmentation Details}
\label{appendix:data_augmentation}
The default pre-training setting, which we also used for our best models, uses 8 audio data augmentations. Not all augmentations are necessarily applied to all inputs: each independent data augmentation is applied with a probability tuned during hyperparameter gridsearch. The most effective augmentations and their probabilities are presented in Section \ref{sec:data_augmentations}. The implementation details for each augmentation are provided below.
We used the `torchaudio-augmentations` Python package for all our experiments \cite{spijkervet_torchaudio_augmentations}.

\subsection{Random Crop}
The audio is cropped with a number of samples $s \in \{20\,736, 43\,740, 59\,049 \}$ for sample rates 8\,000, 16\,000 and 22\,050~Hz respectively. To ensure that every sample in the batch is of the same size, the fragment's window we can crop from with original length $S$ is adjusted to $S-s$.

\subsection{Polarity inversion}
The polarity of the audio signal is inverted by multiplying the amplitude of the signal by $-1$. 

\subsection{Additive White Gaussian Noise}
White Gaussian noise is added to the complete signal with a signal-to-noise ratio (SNR) of 80 decibels.

\subsection{Gain Reduction}
The gain of the audio signal is reduced at random using a value drawn uniformly between -6 and 0 decibels. In our implementation, we use the $\texttt{torchaudio.transforms.Vol}$ interface.

\subsection{Frequency Filter}
A frequency filter is applied to the signal using the $\texttt{essentia}$ library \cite{essentia}. We process the signal with either the $\texttt{LowPass}$ or $\texttt{HighPass}$ algorithm \cite{zolzer2002dafx}, which is determined by a coin flip.

For the low-pass filter, we draw the cut-off frequency from a uniform distributions between 2\,200 and 4\,000~Hz. All frequencies above the drawn cut-off frequency are filtered from the signal.

Similarly for the high-pass filter, we draw the cut-off frequency from a uniform distributions between 200 and 1200~Hz. All frequencies below the cut-off frequency are filtered from the signal.

\newpage

\subsection{Delay}
The signal is delayed by a value chosen randomly between 200 and 500 milliseconds, in 50ms increments. Subsequently, the delayed signal is added to the original signal with a volume factor of 0.5, i.e., we multiply the signal's amplitude by 0.5. An example implementation of this digital signal processing effect is given below in Python using PyTorch:

\begin{verbatim}
    import random
    import torch
    import numpy as np

    ms = random.choice(
        np.arange(200, 500, 50)
    )

    offset = int(ms * (sample_rate / 1000))
    beginning = torch.zeros(audio.shape[0], offset)
    end = audio[:, :-offset]
    delayed_signal = torch.cat((beginning, end), dim=1)
    delayed_signal = delayed_signal * self.volume_factor
    audio = (audio + delayed_signal) / 2
\end{verbatim}

\subsection{Pitch Shift}
The pitch of the signal is shifted up or down, depending on the pitch interval that is drawn from a uniform distribution between -5 and 5 semitones, i.e., up to a perfect fourth higher or lower than the original signal. We assume 12-tone equal temperament tuning that divides a single octave in 12 semitones.

Pitch shifting is done using the $\texttt{libsox}$ library, which is interfaced from the $\texttt{wavaugment}$ Python library \cite{wavaugment2020}.

\subsection{Reverb}
To alter the original signal's acoustics, we apply a Schroeder reverberation effect \cite{schroeder1962natural}. This is again done using the $\texttt{libsox}$ library that is interfaced from the $\texttt{wavaugment}$ Python library \cite{wavaugment2020}.

\section{Additional Experimental Results}
\label{appendix:additional_results}

\subsection{Batch Size}
\label{appendix:exp_mini_batch_size}
The complexity of our contrastive learning approach increases with larger batch sizes, which may result in better representations.
We pre-train from scratch until convergence with varying batch sizes and study its effect on the linear evaluation performance in Table~\ref{tab:mini_batch_ablation}.
While our smallest model already shows competitive performance compared to fully supervised models, the performance increased when using 96 examples per batch.
Our largest model required more parameters to score consistently higher than our middle-sized model ($\approx$25M parameters vs. 2.5M parameters).
We hypothesise that the task of inferring the positive pair of 2.6 second long raw musical audio fragments, in a pool of 254 negative examples ($2\times (128 - 1   )$), may be simply too hard for a smaller encoder.

\begin{table}[h]
    \centering
    \footnotesize
    \begin{tabular}{rcccc}
    \toprule
    &\multicolumn{2}{c}{Tag} &\multicolumn{2}{c}{Clip}\\
    \cmidrule(lr){2-3} \cmidrule(lr){4-5}
    Batch Size & $\text{ROC-AUC}$ & $\text{PR-AUC}$ & $\text{ROC-AUC}$ & $\text{PR-AUC}$  \\
    \midrule
    128 & \textbf{89.7} & \textbf{37.0} & \textbf{94.0} & \textbf{70.0} \\
    96 & 88.7 & 35.6 & 93.0 & 69.2 \\
    48 & 87.9 & 34.6 & 92.9 & 68.8 \\
    \bottomrule
    \end{tabular}
    \caption{Effect of the batch size used during self-supervised training on the linear music classification performance.}
    \label{tab:mini_batch_ablation}
\end{table}

\subsection{Sample Rates}
\label{appendix:sample_rates}
We show in Table~\ref{tab:sample_rate_ablation} that there is a marginal penalty to the final scores for the self-supervised models when re-sampling the audio to 8\,000~Hz and 16\,000~Hz respectively, which is in line with previous work \cite{lee2018samplecnn}.
Since re-sampling disturbs the frequency spectrum, we isolate its contribution by disregarding additional augmentations, i.e., only apply random cropping.

\begin{table}[h]
    \centering
    \footnotesize
      \begin{tabular}{rcccc}
        \toprule
        &\multicolumn{2}{c}{Tag} &\multicolumn{2}{c}{Clip}\\
        \cmidrule(lr){2-3} \cmidrule(lr){4-5}
        SR & $\text{ROC-AUC}$ & $\text{PR-AUC}$ & $\text{ROC-AUC}$ & $\text{PR-AUC}$  \\
        \midrule
        8\,000 & 84.8 & 29.8 & 90.6 & 62.9 \\
        16\,000 & 85.5 & 30.4 & 91.0 & 64.1 \\
        22\,050 & 85.8 & 30.5 & 91.3 & 64.8 \\
        \bottomrule
      \end{tabular}
      \caption{Effect of varying the input audio's sample rate on the linear music classification performance.}
      \label{tab:sample_rate_ablation}
    \end{table}

\subsection{Additional Hidden Layer and Training Duration}
\label{appendix:additional_hidden_layer}
After pre-training with the self-supervised objective, we performed a linear evaluation to test the expressivity of the representations with a classifier of limited capacity. To further assess the representations' usability, we add a single hidden layer to our classifier and again measure the performance on the downstream task of music classification. The results of this experiment are shown in Table \ref{tab:add_hidden_layer} for linear evaluation (left of the forward slashes) as well as when a hidden layer is added (right of the slashes), for different pre-training durations measured in epochs.

Contrastive learning techniques also benefit from longer training compared to their supervised equivalent \cite{chen_simple_2020}.
While larger batch sizes increase the pretext task complexity as shown in Appendix ~\ref{appendix:exp_mini_batch_size}, training longer increases the number of natural variations of the data, which is a diserable goal in representation learning \cite{bengio2013representation}, due to the random augmentation scheme.
We pre-train from scratch until convergence and set the batch size to 96.
Table~\ref{tab:add_hidden_layer} also shows that increasing the self-supervised training duration improves downstream performance.

\begin{table}[h]
    \centering
    \footnotesize
    \begin{tabular}{lcccc}
    \toprule
    &\multicolumn{2}{c}{Tag} &\multicolumn{2}{c}{Clip}\\
    \cmidrule(lr){2-3} \cmidrule(lr){4-5}
    Epochs & $\text{ROC-AUC}$ & $\text{PR-AUC}$ & $\text{ROC-AUC}$ & $\text{PR-AUC}$  \\
    \midrule
    10\,000 & 88.7 / \textbf{89.3} & 35.6 / \textbf{36.0} & 93.2 / \textbf{93.5} & 69.3 / \textbf{70.0} \\
    3\,000 & 88.5 / 88.9 & 35.1 / 35.5 & 93.0 / 93.3 & 69.2 / 69.7 \\
    1\,000 & 88.3 / 88.6 & 34.4 / 34.9 & 92.3 / 93.1 & 68.6 / 69.2 \\ 
    300 & 87.1 / 87.4 & 32.7 / 32.5 & 92.0 / 92.0 & 66.6 / 66.7 \\
    100 & 86.4 / 86.6 & 30.9 / 31.3 & 91.3 / 91.3 & 64.1 / 64.6 \\ 
    \bottomrule
    \end{tabular}
    \caption{Performance difference of a linear classifier and when a single hidden layer is added to the classifier on the downstream music classification performance, for different self-supervised pre-training durations.}
    \label{tab:add_hidden_layer}
\end{table}

\subsection{Qualitative Results}
\label{appendix:qualitative_results}

\begin{figure}[h]
    \centering
    \subcaptionbox{$\boldsymbol{h}_\text{CLMR}$}{\includegraphics[width=.40\textwidth]{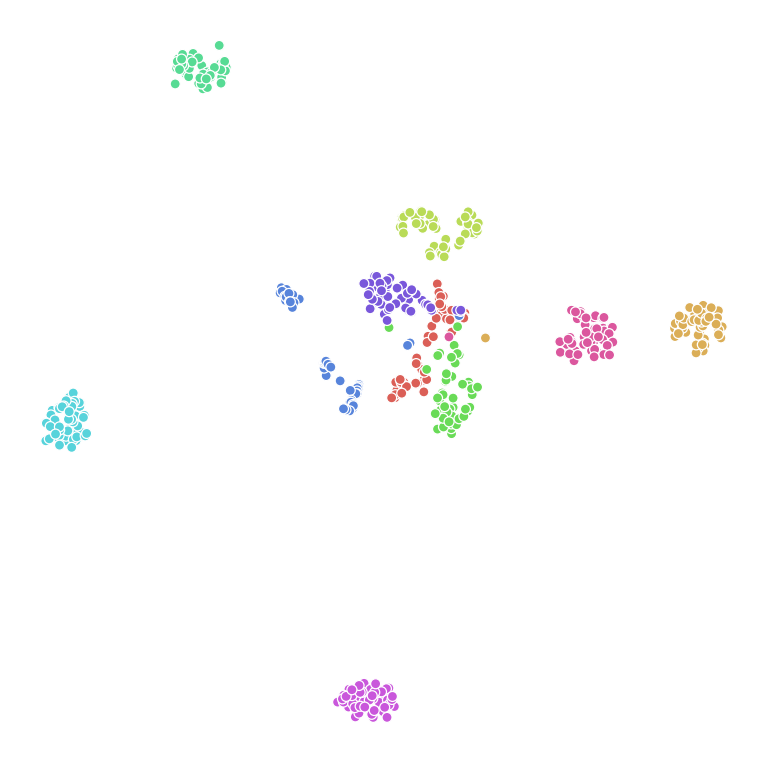}}\hfill
    \subcaptionbox{$\boldsymbol{h}_\text{CPC}$}{\includegraphics[width=.40\textwidth]{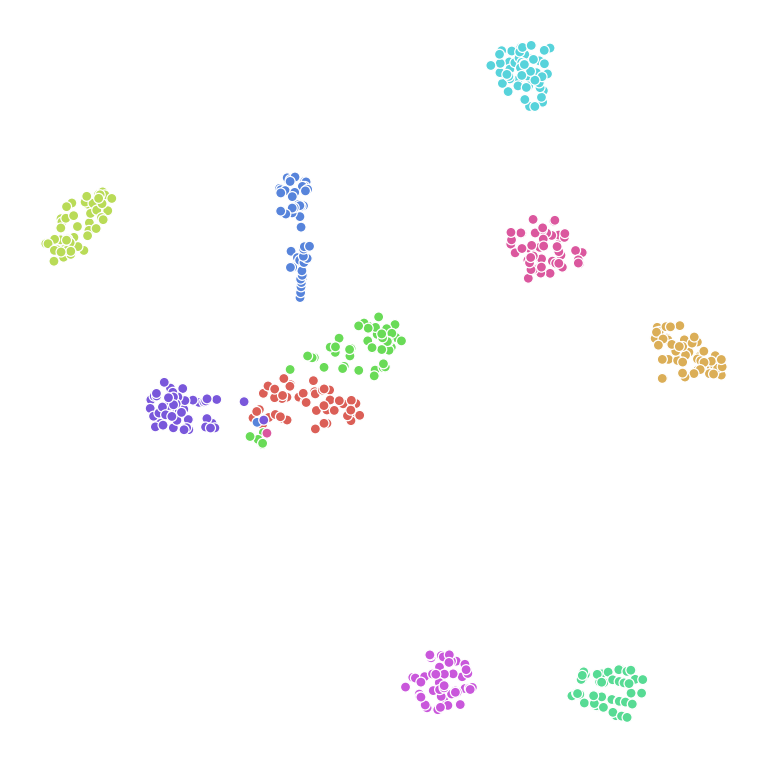}}\hfill
    \caption{$t$-SNE manifolds of the hidden vectors of music audio from a subset of 10 music tracks, i.e., in this case classes, from the validation set. Each point represents a 2.67 second long music fragment belonging to a music track.}
    \label{fig:tsne_manifold}
\end{figure}

Figure \ref{fig:tsne_manifold} shows $t$-SNE visualisations \cite{maaten2008visualizing} of our best self-supervised models representations $\boldsymbol{h}_\text{CLMR}$ and $\boldsymbol{h}_\text{CPC}$, for a randomly set of music tracks from the validation set. We show that both self-supervised models can cleanly seperate the classes.

\begin{figure*}[h]
    \centering
    \subcaptionbox{CLMR$^{(1)}_{\mathrm{MTAT}}$}{\includegraphics[width=.33\textwidth]{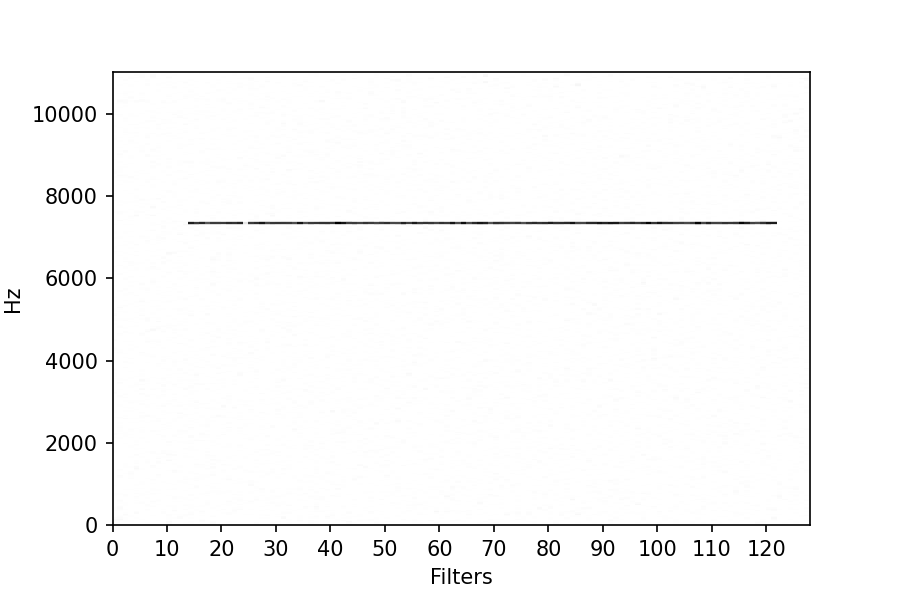}}\hfill
    \subcaptionbox{CLMR$^{(4)}_{\mathrm{MTAT}}$}{\includegraphics[width=.33\textwidth]{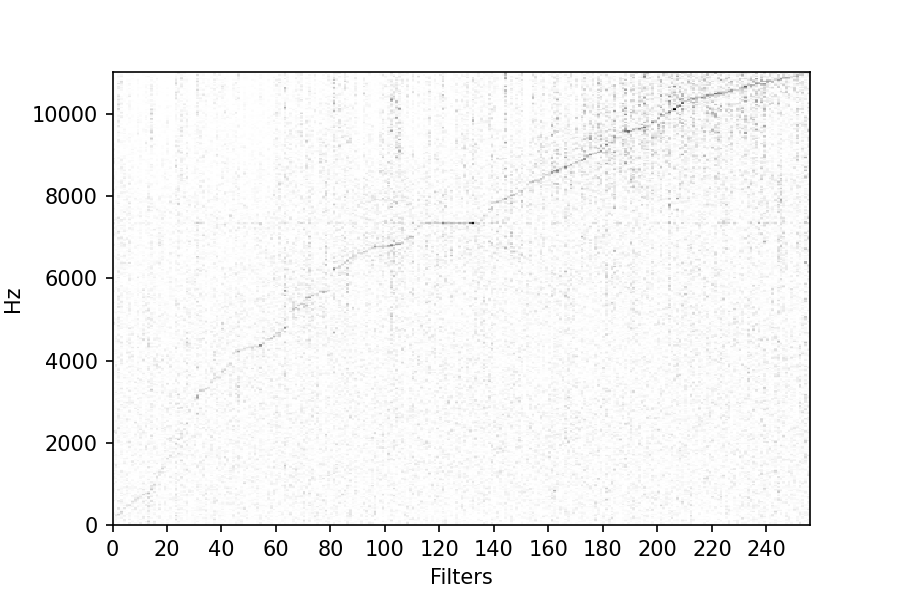}}\hfill
    \subcaptionbox{CLMR$^{(6)}_{\mathrm{MTAT}}$}{\includegraphics[width=.33\textwidth]{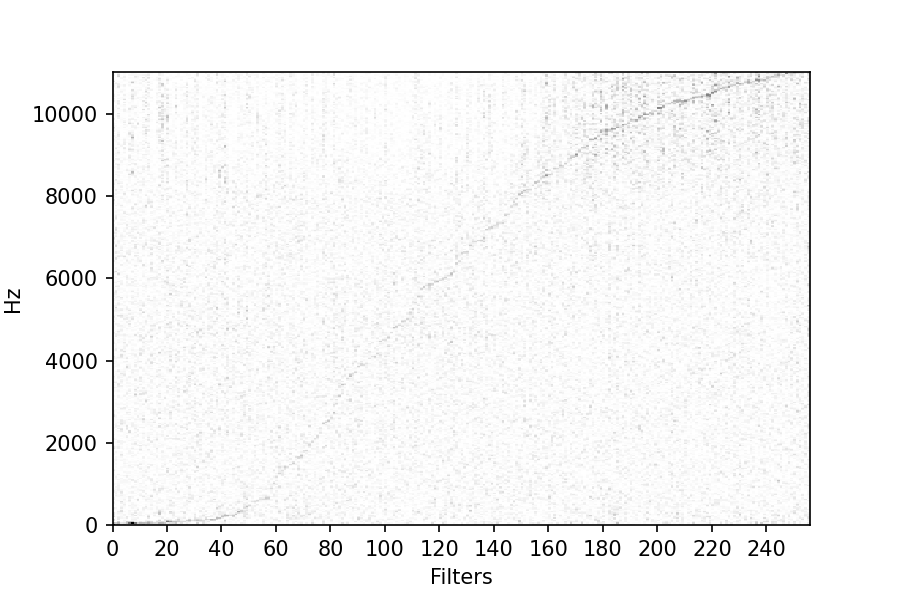}}\hfill
    \subcaptionbox{CPC$^{(1)}_{\mathrm{MTAT}}$}{\includegraphics[width=.33\textwidth]{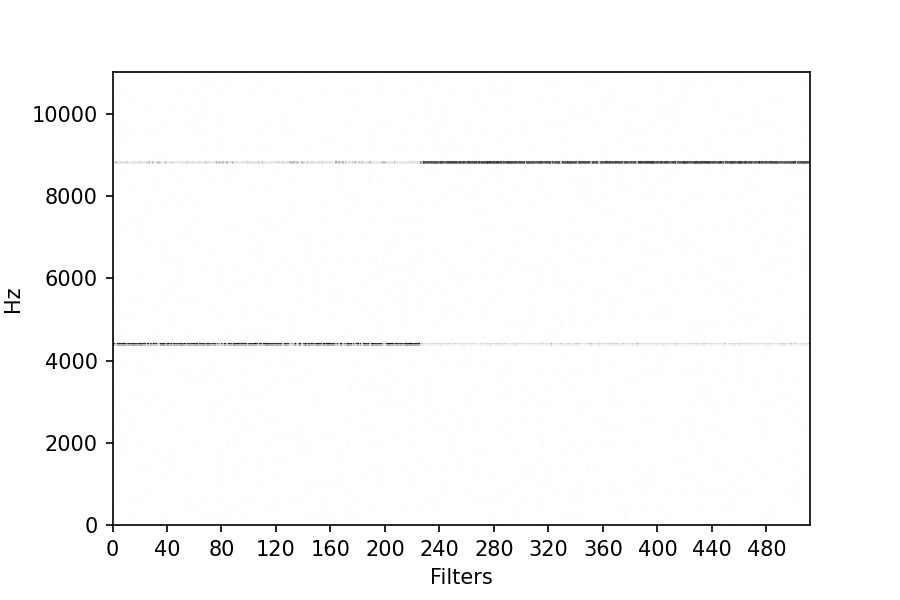}}\hfill
    \subcaptionbox{CPC$^{(4)}_{\mathrm{MTAT}}$}{\includegraphics[width=.33\textwidth]{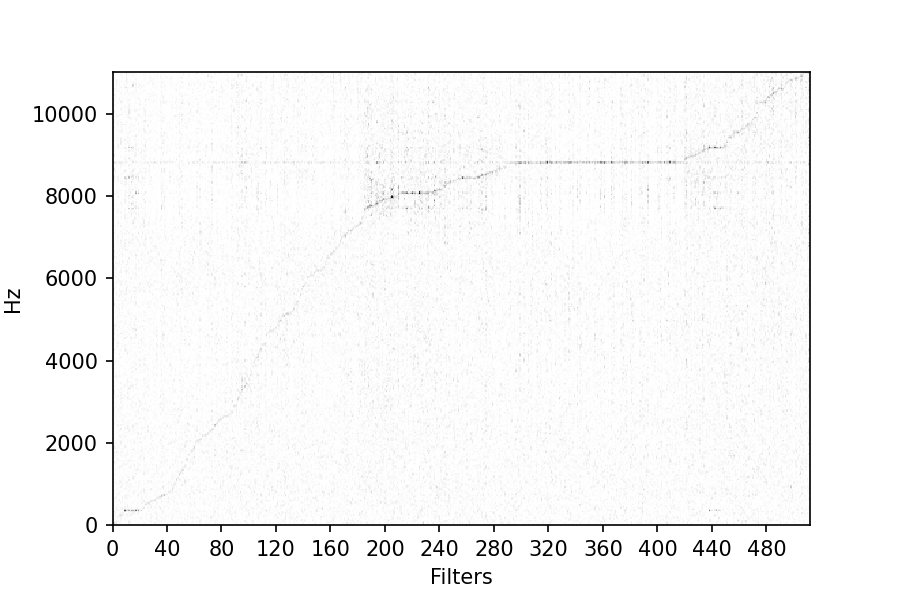}}\hfill
    \subcaptionbox{CPC$^{(6)}_{\mathrm{MTAT}}$}{\includegraphics[width=.33\textwidth]{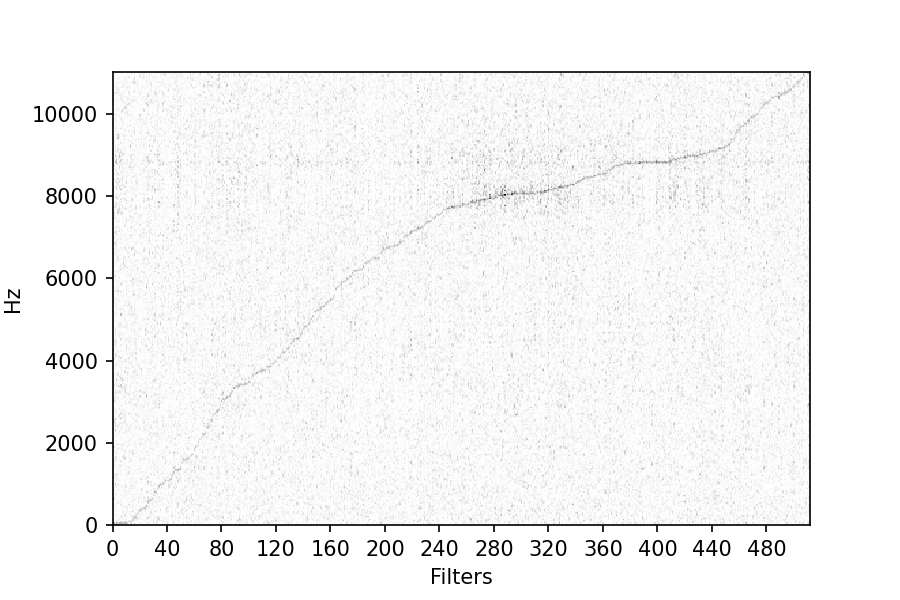}}

    \subcaptionbox{CLMR$^{(1)}_{\mathrm{Billboard}}$}{\includegraphics[width=.33\textwidth]{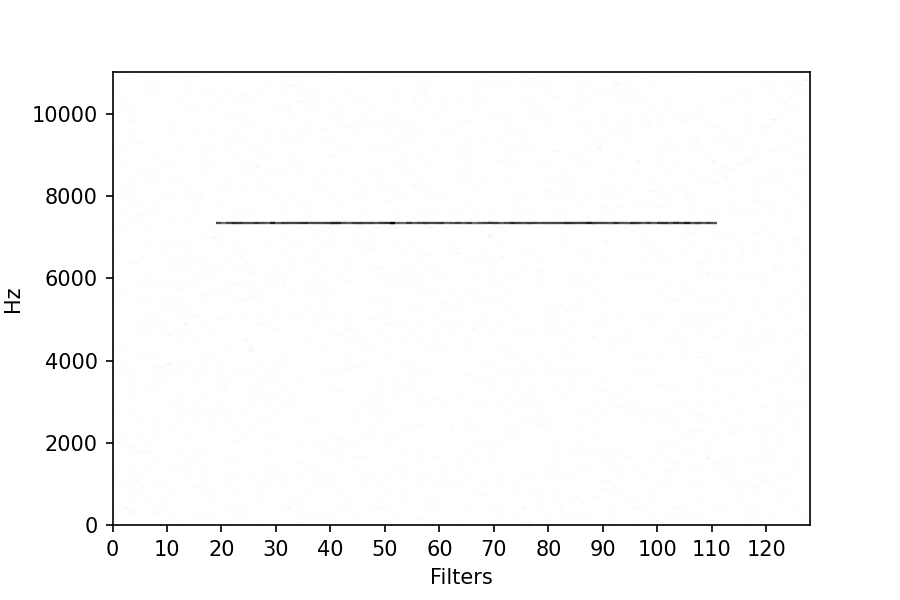}}\hfill
    \subcaptionbox{CLMR$^{(4)}_{\mathrm{Billboard}}$}{\includegraphics[width=.33\textwidth]{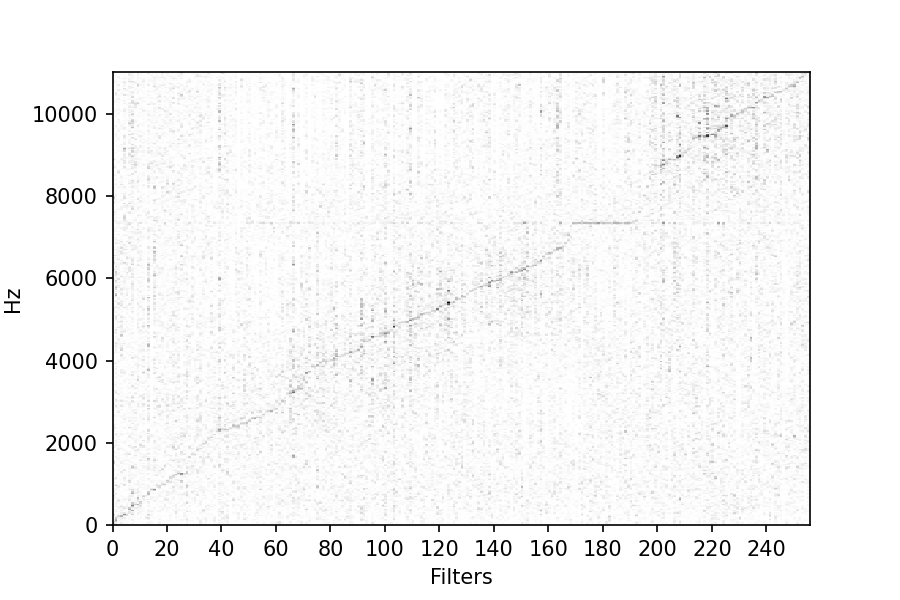}}\hfill
    \subcaptionbox{CLMR$^{(6)}_{\mathrm{Billboard}}$}{\includegraphics[width=.33\textwidth]{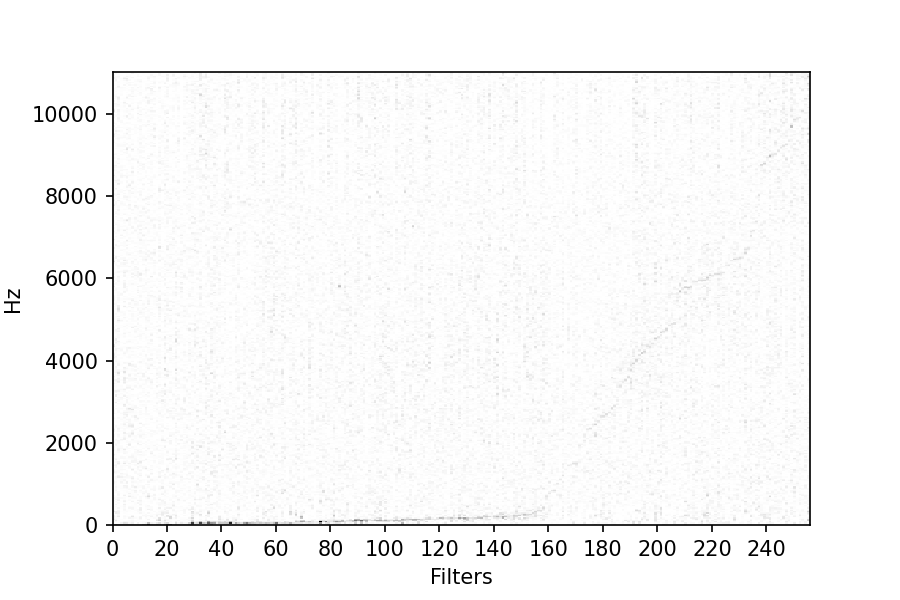}}\hfill
    \subcaptionbox{CPC$^{(1)}_{\mathrm{Billboard}}$}{\includegraphics[width=.33\textwidth]{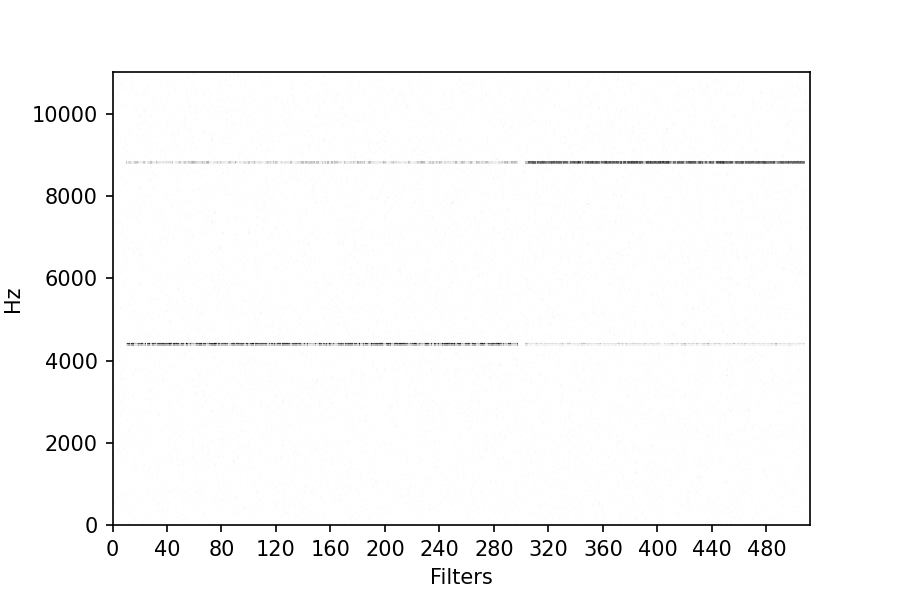}}\hfill
    \subcaptionbox{CPC$^{(4)}_{\mathrm{Billboard}}$}{\includegraphics[width=.33\textwidth]{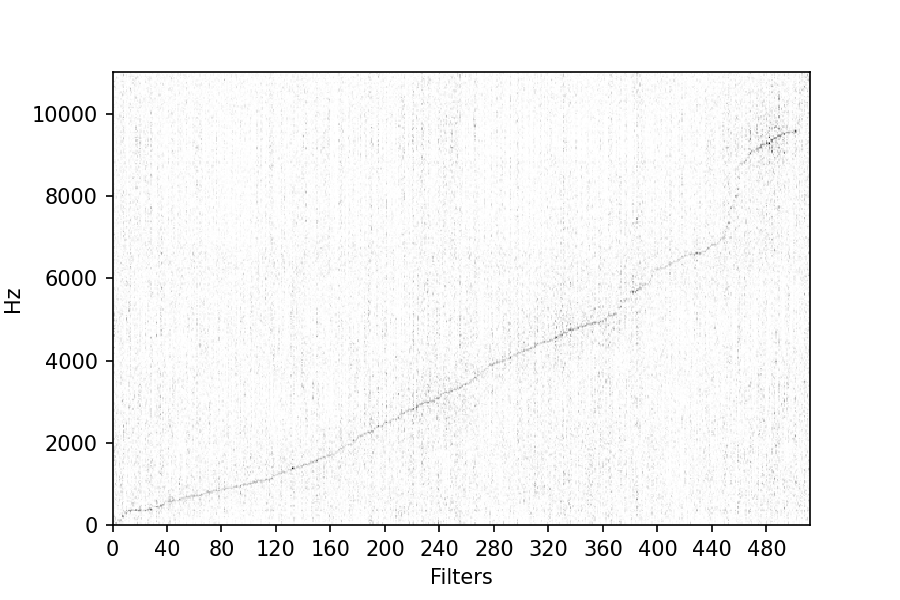}}\hfill
    \subcaptionbox{CPC$^{(6)}_{\mathrm{Billboard}}$}{\includegraphics[width=.33\textwidth]{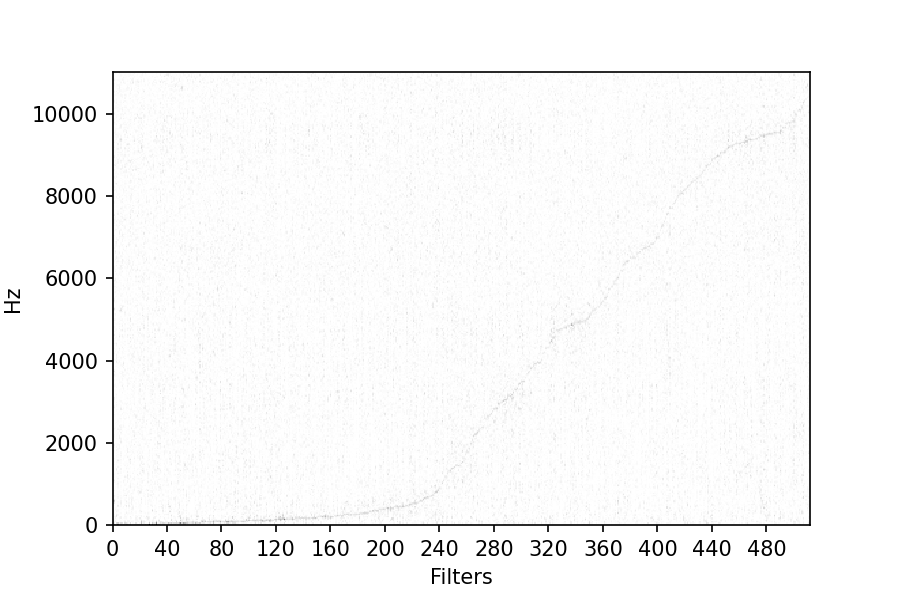}}

    \caption{Normalised magnitude spectrum of the filters of the self-supervised models in the sample-level convolution layers, sorted by the frequency of the peak magnitude.
Gradient ascent is performed on a randomly initialised waveform of 729 samples (close to typical frame size) and its magnitude spectrum is calculated subsequently.
Each vertical line in the graph represents the frequency spectrum of a different filter. The first three images are taken from a pre-trained, converged CLMR model, the last three from a CPC model, on the Magna\-Tag\-A\-Tune or Billboard datasets}
    \label{fig:filter_visualisation}
\end{figure*}

To get an understanding of what the self-supervised models capture from music, we show in Figure~\ref{fig:filter_visualisation} the magnitude spectrum of the learned filters of the sample-level convolutional layers (layers 1, 4 and 6) for CLMR and CPC, pre-trained on the MagnaTagATune dataset. We perform gradient ascent on a randomly initialised waveform of length 729, i.e., a value that is close to a typical frame size and also interacts conveniently with the convolutional structure of the encoder network, and subsequently calculate the magnitude spectrum. The x-axis plots the filter number, the y-axis the magnitude spectrum for a filter number. Lastly, we sort the plot by the frequency of the peak magnitude.

In CLMR, the first layer is sensitive to a single, very small band of frequencies around 7500~Hz, while in higher layers the filters spread themselves first linearly and then non-linearly across the full range.
CPC shows a similar pattern in the higher layers, but shows a strong activation of two frequencies that span an octave in the first layer.
Conversely, the filters of the supervised-trained encoder have a non-linearity that is found in frame-level end-to-end learning \cite{dieleman2014end}, as well as in perceptual pitch scales such as mel or Bark scales \cite{Stevens1937ASF, barkbank_1961}.

\begin{figure*}[h]
    \centering
    \includegraphics[width=\textwidth]{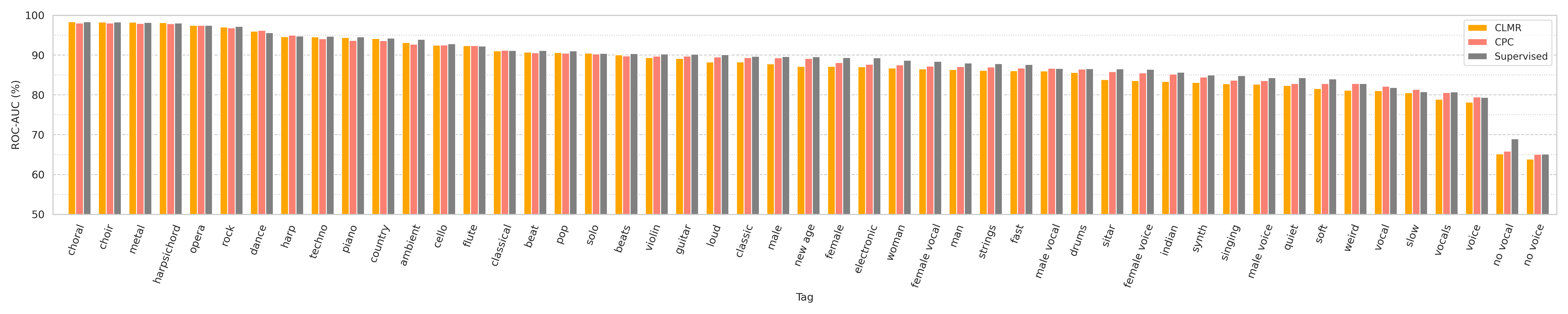}
    \caption{Tag-wise ROC-AUC scores for the top-50 tags in the Magna\-Tag\-A\-Tune dataset, reported for linear, logistic regression classifiers trained on representations of self-supervised models CLMR and CPC, and compared to a fully supervised, end-to-end SampleCNN model.}
    \label{fig:tag_scores}
\end{figure*}

Figure \ref{fig:tag_scores} shows the sorted tag-wise ROC-AUC scores for the top 50 tags in the Magna\-Tag\-A\-Tune dataset, reported for linear evaluation of the trained self-supervised CLMR and CPC models, and the fully end-to-end-trained supervised model. We show that no single tag loses more than 4\% ROC-AUC when trained using self-supervised pre-training and fine-tuning is only performed with a linear classifier, as compared to the supervised benchmark.

\end{document}